# Investigation of the quaternary $Fe_{2-x}Co_xMnSi$ (0 ≤ x ≤ 0.6) alloys by structural, magnetic, resistivity and spin polarization measurements

Lakhan Bainsla,[1,4] K. G. Suresh,[1, #] A. K. Nigam,[2] and M. Manivel Raja,[3] B.S.D.Ch.S. Varaprasad,[4] Y. K. Takahashi,[4] and K. Hono[4]

[1]Department of Physics, Indian Institute of Technology Bombay, Mumbai 400076, India
[2]Tata Institute of Fundamental Research, Homi Bhabha Road, Mumbai 400005, India
[3]Defence Metallurgical Research Laboratory, Hyderabad 500058, India
[4]National Institute for Materials Science, Tsukuba 305-0047, Japan

## Abstract

Detailed study of the $Fe_{2-x}Co_xMnSi$ (0 ≤ x ≤ 0.6) alloys has been carried out by investigating the samples with x-ray diffraction, Mössbauer spectroscopy, magnetization, transport and current spin polarization measurements. A perfectly ordered $L2_1$ phase is found to exist for $x = 0.4$. Competing magnetic interactions between ferromagnetic (FM) and antiferromagnetic (AFM) phases are found to exist in $Fe_{2-x}Co_xMnSi$ for $x < 0.2$ whereas the AFM phase completely disappears for $x \geq 0.2$ as revealed by the magnetization and resistivity data. Anomalous, semiconducting-like behaviour is observed for $x = 0.4$. Current spin polarization has been estimated from different conductance curves obtained by using the point contact Andreev reflection technique. Alloys with x=0.2 and 0.4 show spin polarization values of $0.61 \pm 0.1$ and $0.66 \pm 0.1$ respectively.

---

#corresponding author (#: suresh@phy.iitb.ac.in,)

## I. Introduction

Full Heusler alloys are intermetallic compounds with the composition $X_2YZ$, which crystallizes in the cubic $L2_1$ structure. In these alloys, X and Y atoms are transition metals and Z is a nonmagnetic element. Heusler alloys have recently attracted increasing interest owing to their multifunctional properties[1-5]. Research on half-metallic ferromagnetic materials based on Heusler compounds has been rapidly growing since the prediction of half metallicity in the half Heusler NiMnSb alloy by de Groot et al.[6] Recently, several full Heusler alloys with the cubic $L2_1$ structure and space group Fm3m have also been predicted to be half metallic on the basis of electronic band structure calculations[7]. Since an ideal half metal exhibits 100% intrinsic spin polarization at the Fermi energy $E_F$, these materials have excellent prospects as spin injection electrodes in spintronic devices. Such materials are considered as most potential candidates for the tunneling magnetoresistance devices because the magnetoresistance (MR) is expected to be large in them.

First principle electronic structure calculations have shown that $Fe_2MnSi$ is a half metallic ferromagnet[8]. These calculations have played a key role in predicting several other half metallic ferromagnets as well. It is also known that $Fe_2MnSi$ has an antiferromagnetic phase below $T_R$ (called the re-entry temperature), as reported in our earlier work[9] and also by Booth et al. [10]. With the aim of suppressing the antiferromagnetic phase and to study its effect on the half metallicity, we have partially substituted Co for Fe, resulting in the series $Fe_{2-x}Co_xMnSi$ ($x = 0$, 0.05, 0.1, 0.2, 0.4, 0.6). The Curie temperature ($T_C$) and the saturation magnetization ($M_S$) are found to increase with Co concentration, which is in agreement with the theoretical calculations.[11] High values of spin polarization have been obtained for some compositions such as $x=0.4$, which shows a polarization value of $0.66\pm0.1$.

## II. Experimental Details

Polycrystalline samples of $Fe_{2-x}Co_xMnSi$ alloys ($x = 0$, 0.05, 0.1, 0.2, 0.4, 0.6) were prepared by arc melting under argon atmosphere. The purity of the starting elements was at least 99.9%. The ingots were melted several times to ensure chemical homogeneity and the final weight loss in all the cases has been found to be less than 1%. The melted ingots were annealed under a vacuum of $10^{-5}$ torr for 24 hours at 1073 K, followed by quenching in ice water mixture.

The crystal structure was determined from the X-ray powder diffraction (XRD) data. $^{57}$Fe Mössbauer spectra were recorded at room temperature using a constant acceleration spectrometer with 25 mCi $^{57}$Co(Rh) radioactive source. The obtained spectra were analyzed using PCMOS-II least-squares fitting program. The magnetization (*M*) measurements both under zero field cooled (ZFC) and field-cooled (FC) conditions, in the temperature (*T*) range of 5–400 K and upto a maximum field (*H*) of 50 kOe, were performed by using a vibrating sample magnetometer (VSM) attached to a physical property measurement system (PPMS, Quantum Design). In the ZFC mode, the sample was cooled in the absence of field and at then a field was applied on warming, during which the magnetization was measured. In the FC mode, the sample was cooled in a field, with the cooling field set equal to the measuring field. In this mode the magnetization was measured on both cooling (FCC) and heating (FCW) cycles. Spin polarization measurements were done by using point contact Andreev reflection (PCAR) technique.[12] Sharp Nb tips prepared by electrochemical polishing were used to make superconducting point contacts with the sample. Current spin polarization of the conduction electrons was obtained by fitting the normalized conductance $G(V)/G_n$ curves to the modified Blonder-Tinkham-Klapwijk (BTK) model.[13] A multiple parameter least squares fitting were carried out to deduce spin polarization (*P*) using dimensionless interfacial scattering parameter (*Z*), superconducting band gap ($\triangle$) and *P* as variables. Electrical resistivity was measured in the temperature range of 5–390 K and in fields upto 50 kOe by using the PPMS.

## III.  Results and Discussion

The XRD patterns of the alloys recorded at room temperature are shown in Fig.1. The structural analysis was done with the Rietveld refinement. All the alloys are found to be single phase with cubic L2$_1$ crystal structure and Fm3m (No. 225) space group. The lattice parameter decreases linearly as *x* is increased from 0 to 0.6, though the decrease is only marginal. The linear decrease suggests that this series obeys the Vegard's law[14] and the lattice contraction is generally known to be a desirable trend as far as half metallicity is concerned. This is due to the fact that the shifting of the Fermi level towards the conduction band as a result of lattice contraction, leads to an increase in the band gap in half metals[15].

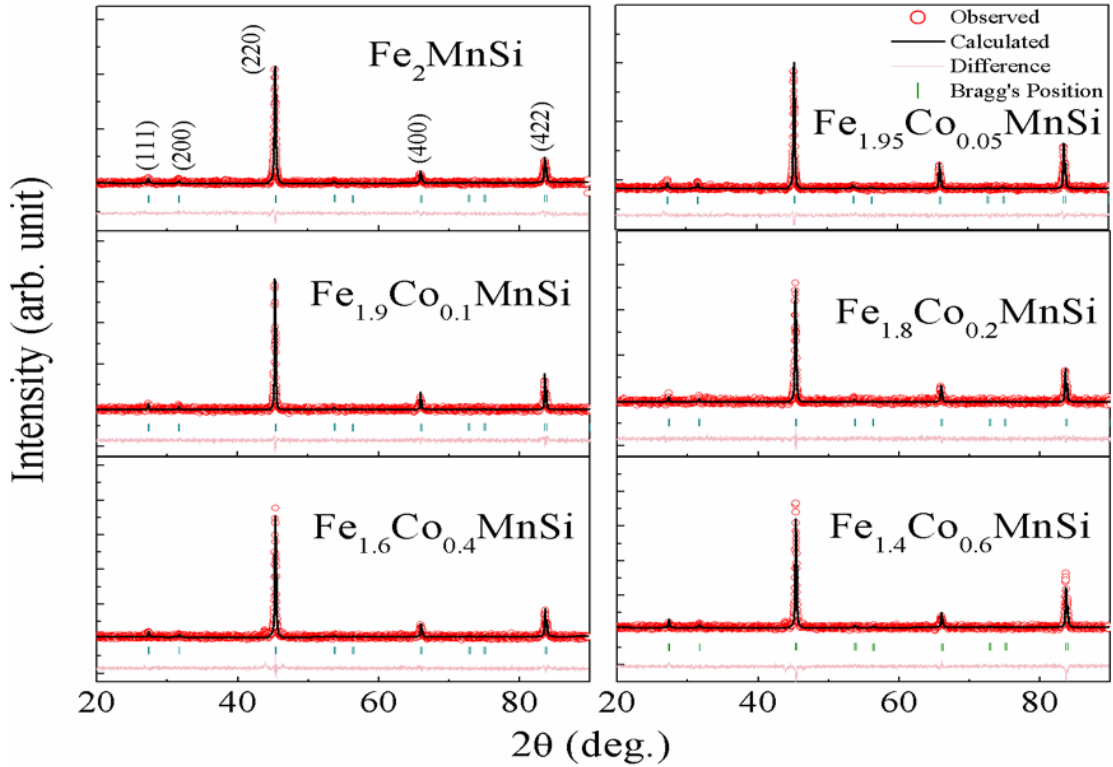

FIG. 1. XRD patterns of $Fe_{2-x}Co_xMnSi$ alloys at the room temperature for $x$ = 0, 0.05, 0.1, 0.2, 0.4, 0.6.

Further study of the crystal structure and ordering has been done by performing $^{57}$Fe Mössbauer spectroscopic (MS) measurements, because perfect $L2_1$ structure is an essential requirement for high spin polarization materials.[16] The parent compound $Fe_2MnSi$ has four lattice sites A, B, C and D. Site A (1/4, 1/4, 1/4)a and and C (3/4, 3/4, 3/4)a correspond to Fe atoms, site B (1/2, 1/2, 1/2)a corresponds Mn atoms and D (0, 0, 0)a for Si atoms. Some amount of atomic disorder has been shown to be present in $Fe_2MnSi$, as reported by $^{57}$Fe Mössbauer spectroscopic[17] and neutron diffraction[10] studies. Therefore, it is of interest to probe this disorder after Co substitution as well. This has been done by doing MS at RT (room temperature) and the collected spectra are shown in fig. 2. The potential of $^{57}$Fe Mössbauer spectroscopy for studying the degree of atomic disorder in iron-containing Heusler alloys has been demonstrated previously for the series $Co_{2-x}Fe_{1+x}Si$[18] and $Co_2Mn_{1-x}Fe_xAl$[19]. To understand the local magnetic environment, $^{57}$Fe Mossbauer spectra were recorded for some alloys ($x$ = 0.1, 0.2, 0.4) at RT by using $^{57}$Co(Rh) source. The fitted Mossbauer parameters such as hyperfine field ($H_{hf}$), quadrupole splitting, isomer shift and the relative intensities of sub-spectra are given in Table 1. The spectra of $x = 0.1$ is fitted with a singlet, which means that the compound is paramagnetic at RT. For $x = 0.2$ there are two sub spectra, a doublet and a sextet with the hyperfine field of 78 kOe, which show ferromagnetically ordered state of $x = 0.2$. However, the presence of doublet reveals that a fraction of Fe is in the paramagnetic state due to near room temperature magnetic transition. Alloy with $x = 0.4$ contains only a sextet, which implies a perfectly $L2_1$ ordered and ferromagnetic state at RT. Moreover, the value of the quadrupole shift is nearly zero which shows cubic symmetry in this alloy. The observed $H_{hf}$ values are comparable to those obtained in earlier studies on $Fe_2$ based alloys[17, 20]. The value of $H_{hf}$ decreases with increase in Co; from 78 kOe to 60 kOe as $x$ increases from $x = 0.2$ to $x = 0.4$. In cubic or nearly cubic compounds, the hyperfine field is determined by the Fermi contact term, which arises due to the polarization of the $s$-electron density at the nucleus. It has been shown that $H_{hf}$ scales linearly with the magnetic moments determined from neutron or magnetization studies[21]. Taking a slope of 125 kOe/$\mu_B$ for Fe germanides from Ref.21, we can roughly estimate the magnetic moment as $0.5\mu_B$ for Fe atoms in the case of $x = 0.4$. For $Fe_2MnSi$, Mn moment is found to be around 2.3$\mu_B$ from the neutron diffraction studies[10]. Considering the total magnetic moment of 3.3$\mu_B$ from the magnetization studies (at 3 K, refer to table 2), the remaining Co moment must be about 0.5$\mu_B$, which is in fair agreement with the calculated values[7].

**Table.1**. Hyperfine field ($H_{hf}$), quadrupole splitting ($\Delta Eq$), isomer shift ($\delta$), line width and relative intensity of sub-spectra of $Fe_{2-x}Co_xMnSi$ alloys at room temperature.

| Alloy | Sub Spectrum | Mossbauer Parameters | | | | |
| --- | --- | --- | --- | --- | --- | --- |
| | | $H_{hf}$ (kOe) | Quadrupole shift ($\Delta E_q$) (mm/s) | Isomer Shift $\delta$ (mm/s) | Line width, (mm/s) | Relative Intensity (%) |
| $Fe_{1.9}Co_{0.1}MnSi$ | Singlet | --- | 0.000 | 0.0973 | 0.31 | 100 |
| $Fe_{1.8}Co_{0.2}MnSi$ | Doublet | --- | 0.774 | 0.0917 | 0.94 | 70 |
| | Sextet | 78 | 0.048 | 0.0814 | 0.96 | 30 |
| $Fe_{1.6}Co_{0.4}MnSi$ | Sextet | 60 | 0.032 | 0.105 | 1.42 | 100 |

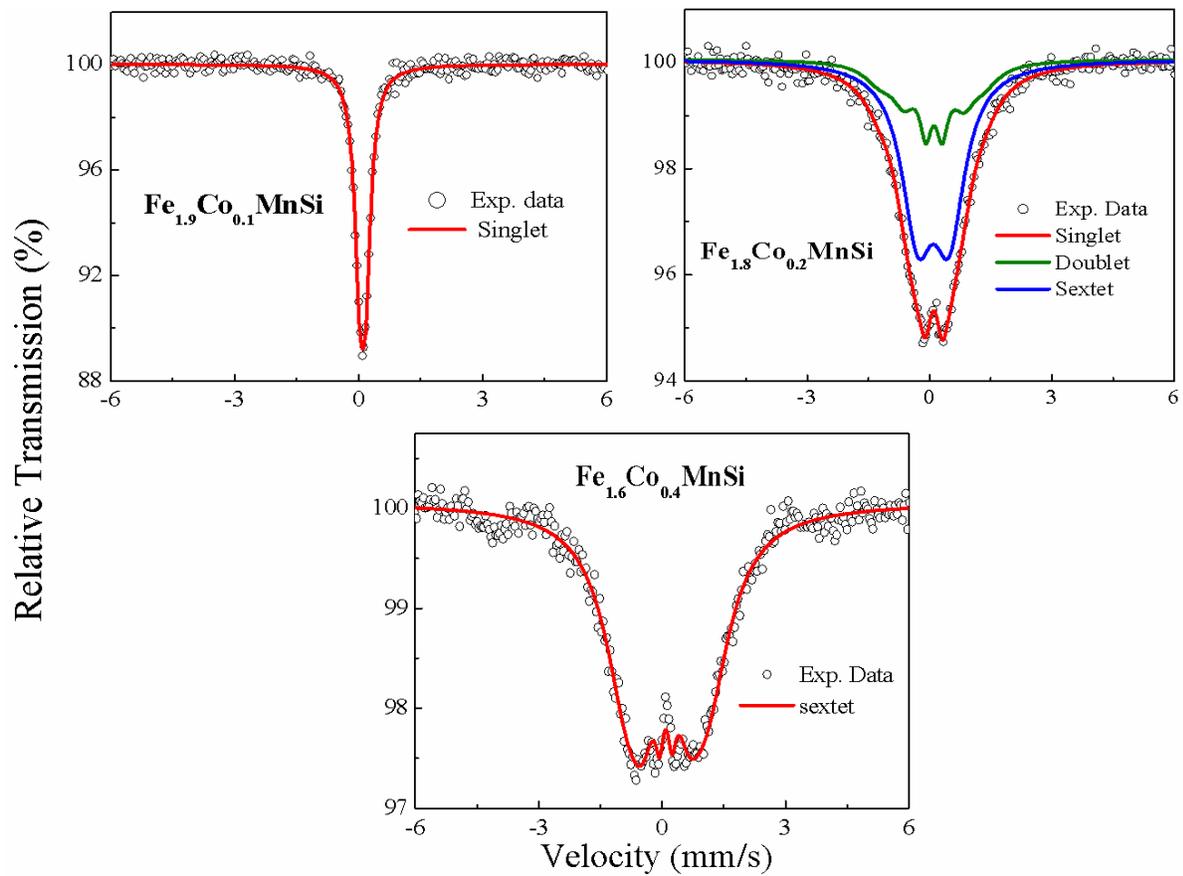

FIG. 2. $^{57}$Fe Mössbauer spectra of Fe$_{2-x}$Co$_x$MnSi ($x$ = 0.1, 0.2, 0.4) alloys collected at room temperature.

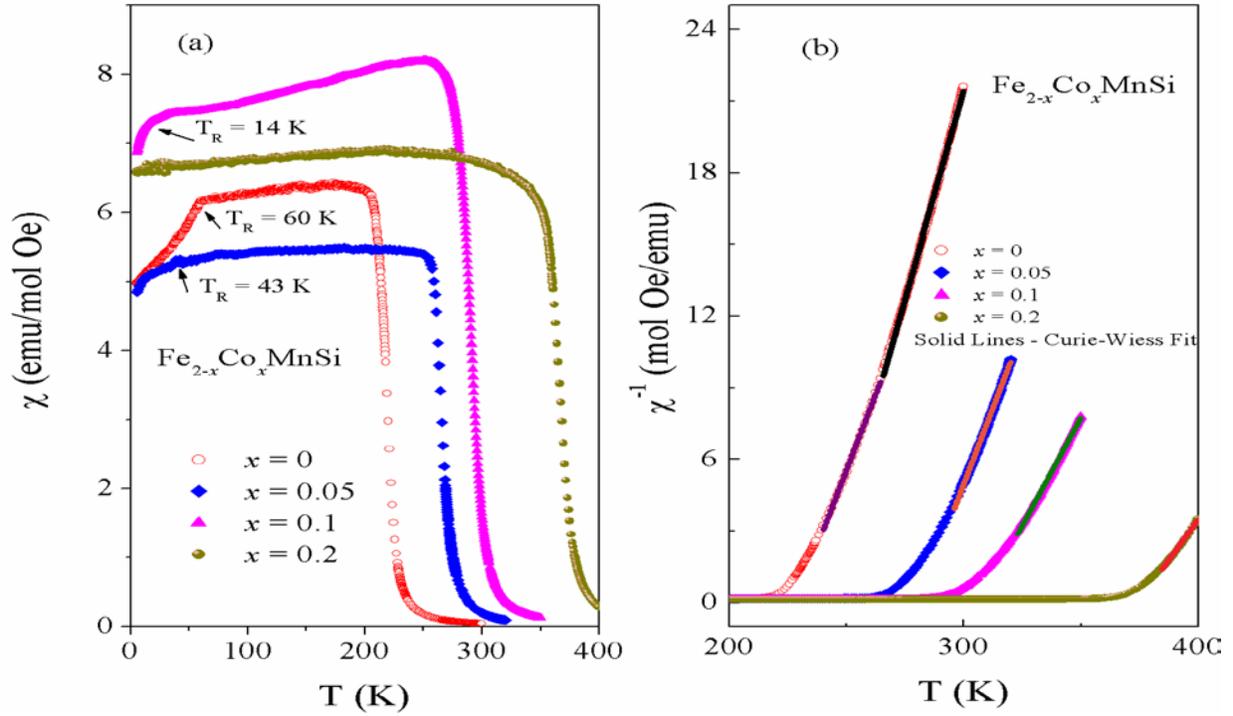

FIG. 3. (a) Temperature dependence of dc magnetic susceptibility for $Fe_{2-x}Co_xMnSi$ ($x$ = 0, 0.05, 0.1, 0.2) in a field of 500 Oe under zero field cooled mode. (b) Inverse susceptibility with the Curie-Weiss fit.

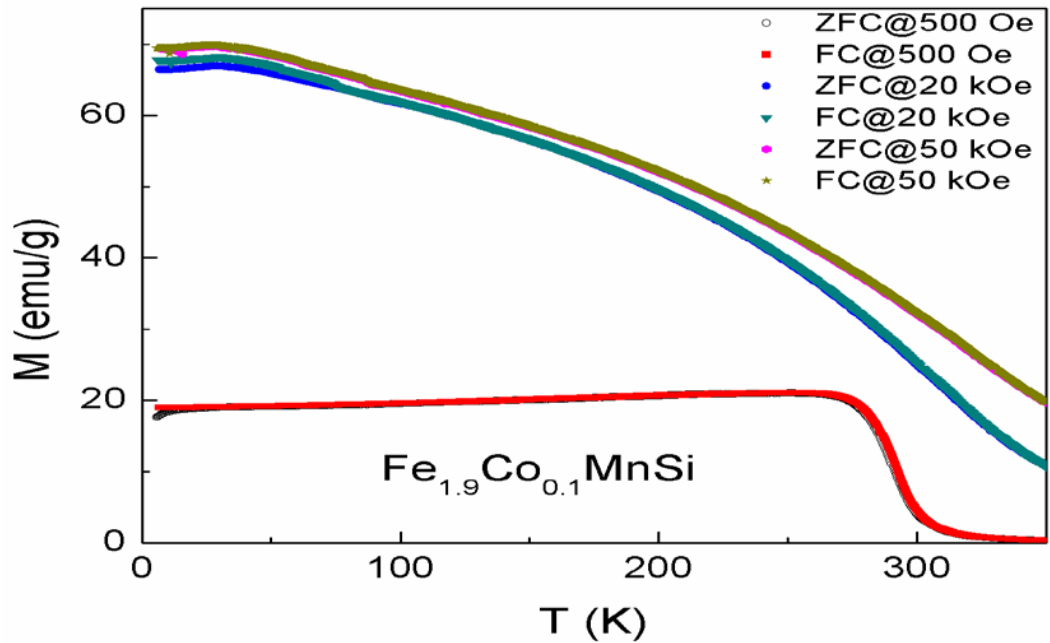

FIG. 4. Thermo magnetic curves for $Fe_{2-x}Co_xMnSi$ alloy with $x$ = 0.1 under different fields

Fig. 3(a) shows the temperature dependence of dc magnetic susceptibility (DCS) measured under ZFC mode in a field of 500 Oe in the temperature range of 5-400 K. These curves show that the Curie temperature increases, while the re-entrant phase transition temperature, $T_R$ decreases with $x$. The re-entrant behaviour completely disappears for $x \geq 0.2$. Curie temperature increases from 220 K for $Fe_2MnSi$ to 368 K for $Fe_{1.8}Co_{0.2}MnSi$. These trends are due the strong hybridization between the d-states of Fe and Co, which enhances the ferromagnetic nature of the alloys as reported theoretically by Ishida et al.[11] The temperature variation of magnetization for $x = 0.1$ under different applied fields is shown in fig. 4, which shows both ZFC and FC data in the temperature range of 5-400 K. There is a bifurcation between the ZFC and the FC plots below the re-entry temperature, as also seen in our earlier work on $Fe_{3-x}Mn_xSi$ and $Fe_2Mn_{1-x}Cr_xSi$ alloys[5,9]. The nature of these curves confirms that the alloy is only weakly antiferromagnetic, unlike the parent $Fe_2MnSi$ alloy.

The variation of the inverse magnetic susceptibility ($\chi$) with temperature and the Curie-Weiss fit are shown in Fig. 3(b). The susceptibility data was fitted by the Curie-Weiss law (solid lines in fig 3(b)). As is evident from the figure, these alloys do not follow the Curie-Weiss behaviour above the transition temperature with a single slope; value of the paramagnetic Curie temperature ($\theta_p$) increases and effective magnetic moments ($\mu_{eff}$) decreases with increase in temperature range of the fit. $\mu_{eff}$ values obtained from the fit in different temperature ranges are listed in table 2. This type of susceptibility behaviour is observed earlier for NiMnSb and PtMnSb as well[22]. An estimate of the Rhodes-Wohlfarth ratio ($p_c/p_s$) suggests that these alloys show itinerant magnetism [23,24]. Here $p_s$ is the saturation moment at low temperatures[25] (3 K in the present case), and $p_c$ is the effective moment per magnetic ion deduced from the Curie constant as given by

$$C_m(emuK/mole) = \frac{N_A p_c (p_c + 2)}{3k_B} \qquad (1)$$

Suppression of antiferromagnetic phase with increasing Co concentration is in good agreement with the theoretical calculations[11], which examined the stability of the ferromagnetic state of $Fe_{2-x}Co_xMnSi$ by comparing the total energy of a few possible antiferromagnetic states vis a vis the ferromagnetic state. According to these calculations, for $x = 0$, one of the assumed antiferromagnetic states is energetically comparable to the ferromagnetic state, which means that the ferromagnetism and the antiferromagnetism are competing with each other. As per these calculations, the ferromagnetic state gets more and more stabilized with increase in *x*, which qualitatively explains the present results.

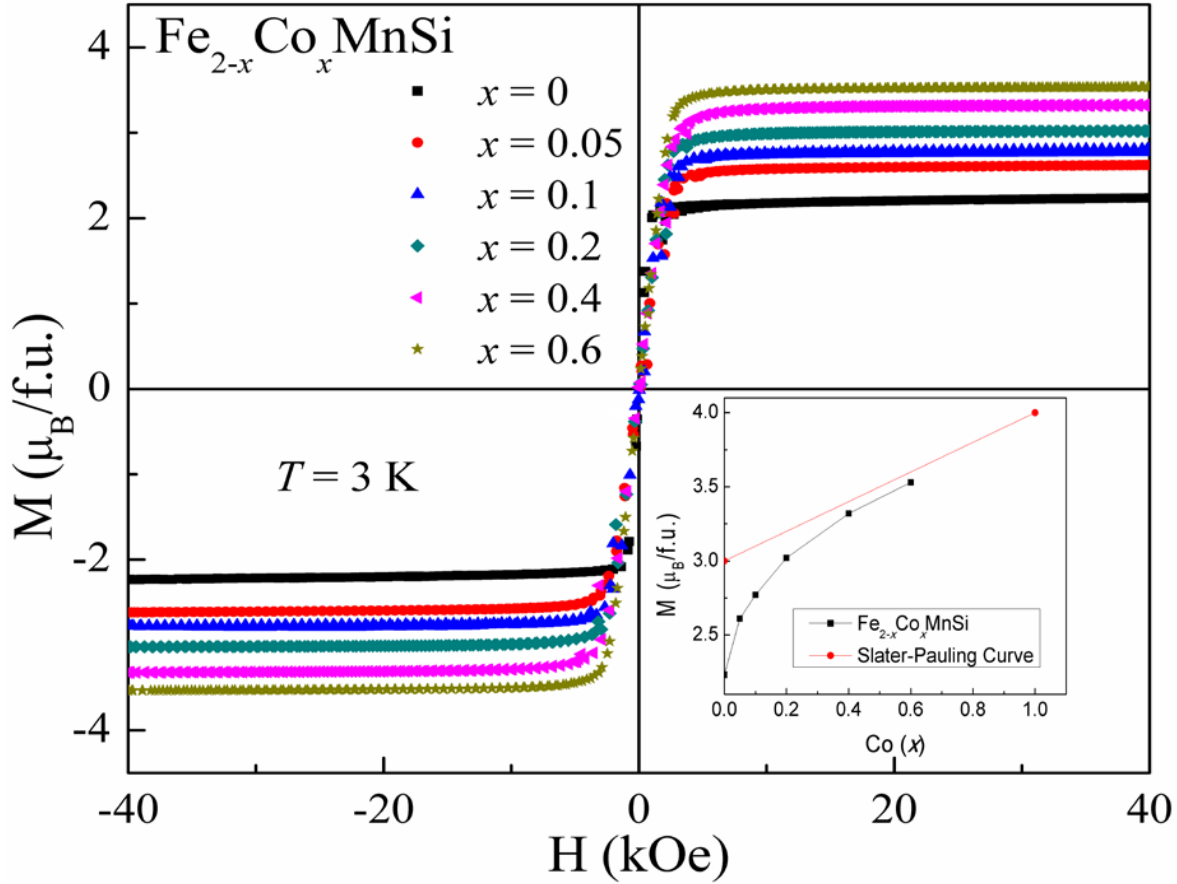

FIG. 5. Isothermal magnetization curves of the $Fe_{2-x}Co_xMnSi$ ($0 \leq x \leq 0.6$) at 3. The inset shows the variation of saturation magnetization with Co concentration along with the expected Slater – Pauling behaviour.

The isothermal magnetization measurements have been done at 3 K in fields up to ±40 kOe and are shown in fig. 5. The alloys show characteristics of soft ferromagnets as the magnetization saturates in low fields. It is also found that the saturation magnetization increases with *x*. $M_S$ value increases from 2.2 $\mu_B$/f.u. for $Fe_2MnSi$ to 3.5 $\mu_B$/f.u for $Fe_{1.4}Co_{0.6}MnSi$. Ishida et al.[11] have stated that the strong hybridization between the *d* states of the Fe and Co atoms, which affects the local density of states (DOS) of Fe near the Fermi level is responsible for this behaviour. The filling of the DOS of Fe increases for the up spin sub-band with increasing $x$, while the moments on the Co and Mn atoms are nearly constant because the local DOS is mostly occupied in the up spin state. Because of this, the moment on Fe increases linearly with $x$, causing the increase in the total moment. Inset of fig.5 shows the variation of saturation magnetization with Co concentration. As can be seen, at high Co concentrations, our results for saturation magnetization closely match with those expected on the basis of the Slater-Pauling rule proposed for full Heusler alloys[26].

**Table. 2** Crystallographic and magnetic properties of $Fe_{2-x}Co_xMnSi$ alloys ($x$ = 0, 0.05, 0.1, 0.2, 0.4, 0.6).

| Compound | $a$ (Å) | $T_R$ (K) | $T_C$ (K) | $M_S^{3K}$ ($\mu_B$/f.u.) | $\theta_p$ (K) | $\mu_{eff}$ ($\mu_B$) | $p_c/p_s$ |
|---|---|---|---|---|---|---|---|
| $x = 0$ | 5.664 | 60 | 220 | 2.2 | 228-238 | 5.6-4.8 | 2.1-1.8 |
| $x = 0.05$ | 5.663 | 43 | 264 | 2.6 | 280 | 5.5 | 1.8 |
| $x = 0.1$ | 5.663 | 14 | 298 | 2.7 | 307 | 6.6 | 2.1 |
| $x = 0.2$ | 5.661 | - | 368 | 3.0 | 374 | 7.7 | 2.25 |
| $x = 0.4$ | 5.660 | - | - | 3.3 | - | - | - |
| $x = 0.6$ | 5.658 | - | - | 3.5 | - | - | - |

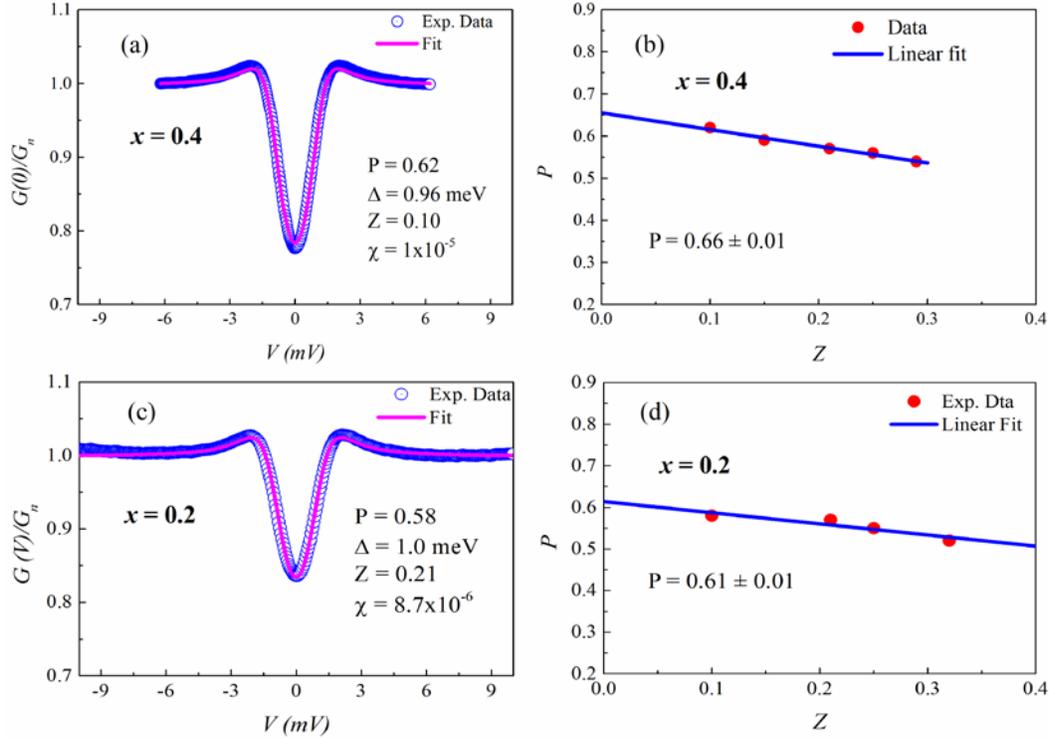

FIG. 6. Typical normalized conductance curves with low value of interfacial scattering amplitude (Z) measured at 4.2 K, In fig. (a) and (c) open circles denote the experimental data and solid lines are the fit to the data by using modified BTK model. Figures (b) and (d) P vs. Z curves for x = 0.4 and 0.2 respectively.

Spin polarization measurements have been performed for few compositions by using PCAR technique. The measurements have been performed on $x = 0.2$ and $x = 0.4$ compositions only, because other compositions have low $T_C$ and hence not interesting from the point of view of applications at RT. The alloys with $x = 0.2$ and 0.4 have high Curie temperatures and $x = 0.4$ is found to possess highly ordered L2$_1$ structure. Typical normalized conductance curves obtained for these compositions are shown in fig. 6. Experimental data is fitted according to modified BTK model[13] using spin polarization, superconducting gap and interfacial scattering parameter as variables. Here we have shown the data collected at low Z values. The shape of the conductance curves near the superconducting gap depends on the value of Z; for low values of Z, the curves are more flat near Δ. The values of various parameters after fitting are shown in the figure. Δ values are small compared to those of the bulk superconducting gap value (1.5 meV), which may be the result of multiple contacts[27]. The intrinsic spin polarization value is ideally obtained by achieving $Z = 0$. But in our case the lowest values are $Z = 0.10$, resulting in $P = 0.62$ and $P = 0.58$ for $x = 0.2$ and $x = 0.4$ respectively. Therefore, the intrinsic value of spin polarization was estimated by linear fitting of the $P$ vs. $Z$ data with extrapolation down to $Z = 0$. This yielded intrinsic spin polarization values of $0.61 \pm 0.1$ and $0.66 \pm 0.1$ for $x = 0.2$ and $x = 0.4$ respectively. The estimated spin polarization value for $x = 0.4$ is higher than that of many ternary or quaternary Heusler alloys. [28-30]

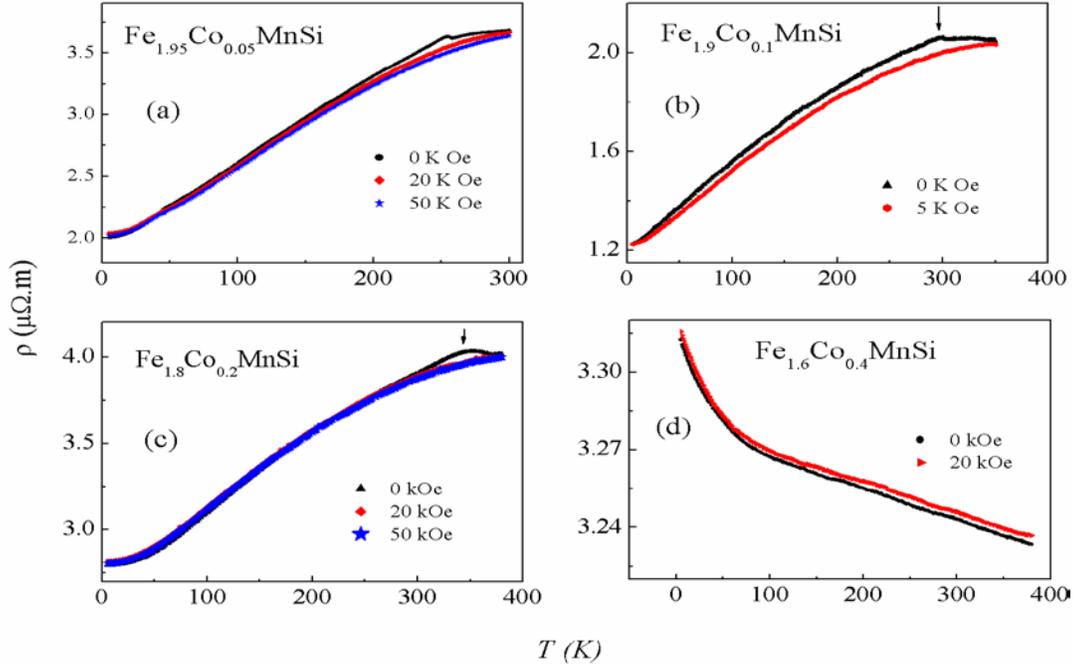

FIG. 7. Temperature variation of electrical resistivity in different fields for the $Fe_{2-x}Co_xMnSi$ ($x = 0.05, 0.1, 0.2$ and $0.4$) compounds.

The temperature variation of the electrical resistivity was measured for the compositions $x = 0.05, 0.1, 0.2$ and $0.4$ in fields of 0, 20, 50 kOe and are shown in fig. 7. There is a clear change of slope near the transition temperatures ($T_R$ and $T_C$), which is due to the reduction in the spin fluctuations with the field, near transition temperatures.

It is clear that the metallic nature changes to semiconducting-like as $x$ is increased to 0.4. Alloys with $x = 0.05, 0.1$ and $0.2$ have metallic behaviour up to the Curie temperature. For $x = 0.1$ and $0.2$ resistivity shows a negative temperature coefficient above the Curie temperature and follows semiconducting-like behaviour, while in x=0.05, it is almost a constant above $T_C$. This type of anomaly above the transition temperature has been observed earlier in $Fe_2Mn_{1-x}Cr_xSi$[5] and other compounds.[31] On the other hand, for $x = 0.4$, resistivity follows a semiconducting-like behaviour throughout the entire temperature regime investigated. Such a resistivity behaviour has been reported for some other Heusler alloys as well. For example, Nishino et al. have studied the transport properties of $Fe_2VAl$[31] and Quardi et al. have recently reported the spin gapless behaviour of $Mn_2CoAl$ by resistivity and Hall effect measurements.[32] $Mn_2CoAl$ shows high value of resistivity, low charge carrier concentration and low Seebeck coefficient, which demonstrates the stability of electronic structure and insensitivity to the disorder.

The anomalous resistivity behavior of $x$=0.4 alloy can be explained by taking into account the Mooij criterion. According to Mooij criterion[33], high value of electrical resistivity in metallic conductors generally tends to reduce the temperature coefficient of resistivity, finally making it negative above a critical value of resistivity. It has been reported by Isino and Muto[34] that this type of resistivity behavior is independent of the electronic structure.

Among the alloys of the present series, $Fe_{1.6}Co_{0.4}MnSi$ is found to have a high Curie temperature and high value of saturation magnetization, along with a negative temperature coefficient of resistivity. Moreover, this alloy is fully ordered as verified by Mössbauer spectroscopy (data is fitted by single sextet) and therefore, the linear behavior cannot be attributed to the disorder. Such a behavior of resistivity is also reported for $Mn_2CoAl$[32] and $NiMnSb$[35]. It may be noted that the resistivity values of $Mn_2CoAl$ and $PtMnSn$ are $\approx 400$ µΩ-m and $\approx 0.5$ µΩ-m, respectively. Hence the alloy with $x = 0.4$ shows some interesting transport properties.

According to Mattheissen's rule[36], the total resistivity of a ferromagnetic material is the sum of the residual resistivity, the phonon contribution ($\rho_{ph}$) and the magnon contribution ($\rho_{mag}$). The resistivity data of x=0.05 and 0.1 alloys show that the electron-phonon interaction dominates in the high temperature region, which indicates the presence of band gap between the spin up and spin down states at the Fermi level (electron-magnon interactions are very small compared to electron-phonon interactions). In the low temperature region, $x = 0.1$ shows a clear absence of $T^2$ dependence, indicating a sign of half metallicity. Indirect evidence of half metallicity based on the absence of $T^2$ term has been reported in some other Heusler alloys.[5,9] It is found that the value of the residual resistivity in zero field is lower than that with field for $x = 0.05$. But such a difference does not exist for the $x \geq 0.2$. This may be due to the complex magnetic phase below $T_R$, due to which the antiferromagnetic spin fluctuations are induced by the applied field in the case of $x=0.05$.

The magnetoresistance, defined as the percentage change in the resistivity[37] has been calculated and the field dependence of MR for the compounds with $x = 0.05$ and 0.2 is shown in Fig.8. In the case of $x = 0$ (Ref. 9), the MR is found to be nominally negative in the regime $T_R < T < T_C$, while it is very small and almost field independent in the paramagnetic region. Below $T_R$, the MR is nominally positive. This must be due to the spin fluctuations induced by the field on the canted spin structure that causes an increase in the resistivity. Also the MR curves show a strong field induced irreversibility in the antiferromagnetic region (at 5 K). These trends are almost retained in $x=0.05$ as well, as can be seen from the figure. The zero field value at the beginning of the first field increasing path (path 1) is quite different from the same as the field is cycled in the decreasing path (path 2), indicating the field-induced irreversibility. Interestingly, for $x = 0.2$, there is no such field induced irreversibility, which did not show the re-entrant phase. This clearly shows that canted moments of Mn align ferromagnetically during path 1 and with the removal of field (path 2), the spins do not come to their original canted structure. It may also be noted that the compound with $x = 0.05$ shows the maximum negative MR of about 3.5%, and for $x = 0.2$, it is about 2.5% near the Curie temperature; observed MR values are in close agreement with those predicted for $Fe_{3-x}Mn_xSi$ alloys.[9]

Presence of field-induced irreversibility of MR at 5 K (antiferromagnetic region) may be due to the two competing magnetic phases (ferromagnetic component along with the antiferromagnetic phase) and the development of a non-cubic phase[26,38]. The irreversibility seen in this case is similar to that seen in $Nd_5Ge_3$[39]. It is found that the presence of this irreversibility is closely related to the occurrence of reentrant antiferromagnetic phase. The canted spin structure of $Fe_2MnSi$ may be responsible for this kind of scenario.[10,40-44] Therefore, one can see a direct correlation between the structural, magnetic and transport properties in this system.

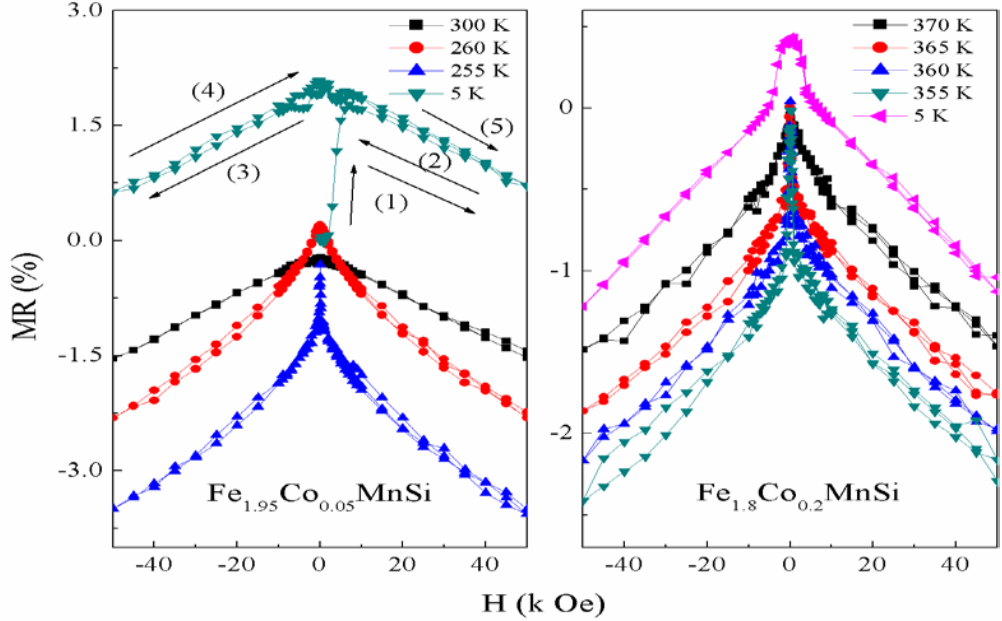

FIG. 8. Magneto-resistance curves of $Fe_{2-x}Co_xMnSi$ ($x$ = 0.05, 0.1) alloys.

## IV. Conclusions

Pseudo quaternary $Fe_{2-x}Co_xMnSi$ with various $x$ values have been studied in detail by structural, magnetic, transport and spin polarization measurements. A highly ordered $L2_1$ phase is observed for $x$ = 0.4 from the x-ray diffraction and Mössbauer data. Substitution of Co in place of Fe in $Fe_2MnSi$ suppresses the antiferromagnetic phase and stabilizes the ferromagnetic state as evident from magnetization and resistivity data. High value of spin polarization of 0.66±.1 is estimated for $x$ = 0.4. Resistivity measurements give indirect evidence of half metallicity in the alloy with x=0.1. Considering the high values of spin polarization and $T_C$, composition with $x$ = 0.4 appears quite promising for applications.

## Acknowledgements

Lakhan Bainsla would like to thank UGC, New Delhi for granting fellowship. The authors thank D. Buddhikot for his help in the resistivity measurements. KGS thanks ISRO cell, IITB for funding the work.

# References

[1] T. Graf, C. Felser, and S. S. P. Parkin, Prog. Solid State Chem. **39**, 1 (2011).

[2] Y.V. Kudryavtsev, N.V. Uvarov, V.N. Iermolenko, I.N. Glavatskyy and J. Dubowik, Acta materialia 60 (2012) 4780-4786.

[3] T. Samanta, I. Dubenko, A.l Quetz, S. Stadler, and N. Ali, Appl. Phys. Letter 101, 242405 (2012).

[4] A. K. Nayak, M. Nicklas, S. Chadov, C. Shekhar, Y. Skourski, J. Winterlik, and C. Felser, Phys. Rev. Letter 110, 127204 (2013).

[5] L. Pal, Sachin Gupta, K. G. Suresh and A. K. Nigam, J. Appl. phys. **115**, 17C303 (2014).

[6] R.A.deGroot, F.M.Müller, P. G. van Engen, andK.H. J.Buschow, Phys. Rev. Lett. **50**, 2024 (1983).

[7] J. Kübler, A. R. Williams, and C. B. Sommers, Phys. Rev. B **28**, 1745 (1983).

[8] M. Lezaic, P. Mavropoulos, and S. Blugel, Hubert Ebert, Phys. Rev. B **83**, 094434 (2011).

[9] L. Pal, K. G. Suresh and A. K. Nigam, J. Appl. Phys. 113, 093904 (2013).

[10] S. Yoon and J. G. Booth, J. Phys. F **7**, 1079 (1977).

[11] S. Ishida, D. Nagatomo, S. Fujji and S. Asano, Materials Trans. Vol. 49 No. 1 (2008) pp 114 119.

[12] R. J. Soulen, J. M. Byers, M. S. Osofsky, B. Nadgorny, T. Ambrose, S. F. Cheng, P. R. Broussard, C. T. Tanaka, J. Nowak, J. S. Moodera, A. Barry, J. M. D. Coey, Science 25, 282 (1998).

[13] G. J. Strijkers, Y. Ji, F. Y. Yang, C. L. Chien, J. M. Byers, Phys Rev B 63, 104510 (2001).

[14] A. R. Denton and N. W. Ashcroft, Phy. Rev. A, 43 (1991) 3161-3164.

[15] I. Galanakis, Ph mavropoulos, P. H. Dederichs, J. Phys. D: Appl. Phys. 39, 765 (2006).

[16] C. Felser, G. H. Fecher, and B. Balke, Angew. Chem., Int. Ed. **46**, 668 (2007).

[17] S. H. Mahmood, A. F. Lehlooh, A. S. Saleh and F. E. Wagner, Phys. stat. sol. (b) 241, No. 6, 1186–1191 (2004).

[18] V. Jung, B. Balke, G. H. Fecher, and C. Felser, Appl. Phys. Lett. **93**, 042507 (2008).

[19] V. Jung, G. H. Fecher, B. Balke, V. Ksenofontov and C. Felser, J. Phys. D: Appl. Phys. **42**, 084007 (2009).

[20]Ghassan A. Al-Nawashi, Sami H. Mahmood, Abdel-Fatah D. Lehlooh and Ahmad S. Saleh, Physica B 321 (2002) 167–172.

[21]O. Massenet, H. Daver, V. D. Nguyen, and J. P. Rebouillat, J. Phys. F: Met. Phys. **9**, 1687 (1979.

[22]Otto M J, Van Woerden R **A** M, Vander Valk P J, Wijngaardl J, Van Bruggent C F, Haas C and Buschow K H J, J. Phys.: Condens. Matter l(1989) 2341-2350.

[23]E. P. Wohlfarth, J. Magn. Magn. Mater. 7 (1978), 113-120.

[24]P. Rhodes, E. P. Wohlfarth, Proc. R. Soc. A 273 (1963) 247.

[25]N. H. Dung, L. Zhang, Z. Q. Ou, L. Zhao, L. Van Eijck, A. M. Mulders, M. Avdeev, E Suard, N. H. Van Dijk and Bruck Ekkes, Phys. Rev. B 86, 045134 (2012).

[26]I. Galanakis, P. H. Dederichs, N. Papanikolaou, Phys. Rev. B 66, 174429 (2002).

[27]S. K. Clowes, Y. Miyoshi, O. Johannson, B. J. Hickey, C. H. Marrows, M. Blamire, M. R. Branford, Y. V. Bugoslavsky, L. F. Cohen, J. Magn. Magn. Mater. 272, 1471 (2004).

[28]S.V. Karthik, A. Rajanikanth, Y.K. Takahashi, T. Ohkubo, K. Hono, Acta Materialia 55, 3867–3874 (2007).

[29]T. M. Nakatani, A. Rajanikanth, Z. Gercsi, Y. K. Takahashi, K. Inomata and K. Hono, J. Appl. Phys. **102**, 033916 (2007).

[30]A. Rajanikanth, Y. K. Takahashi and K. Hono, J. Appl. Phys. **101**, 023901 (2007).

[31]Y. Nishino, M. Kato, S. Asano, K. Soda, M. Hayasaki, and U. Mizutani, Phys. Rev. Lett. VOLUME 79, NUMBER 10, 1909-1912 (1997).

[32]Siham Ouardi, Gerhard H. Fecher, Claudia Felser and Jurgen Kubler, Phys. Rev. Lett. 110, 100401 (2013).

[33]J. H. Mooij, Phys. Status Solidi A 17, 521 (1973).

[34]M. Isino and Y. Muto, J. Phys. Soc. Jpn. 54, 3839 (1985).

[35]J. S. Moodera and D. M. Mootoo, J. Appl. Phys. 76, 6101 (1994).

[36]A. Kowalczyk, L. Smartz, Journal of alloys and compounds 259 (1997) 59-61.

[37]Sachin B. Gupta, K. G. Suresh, A. K. Nigam, J. Appl. Phys. 112, 103909 (2012).

[38]S. Yoon and J. G. Booth J G, J. Phys. Lett. Volume 48A, Number 5 381 (1974).

[39]B. Maji, K. G. Suresh and Nigam A K, *Euro Phys. Lett.* **91** 37007 (2010).

[40]V. Niculescu, T. J. Burch, and J. I. Budnick, J. Magn. Magn. Mater. **39**, 223 (1983).

[41]V. Niculescu, K. Raj, T. J. Burch, and J. I. Budnick, Phys. Rev. B **13**, 3167 (1976).

[42]S. Yoon and J. G. Booth, Phys. Lett. **48** A, 381 (1974).

[43]P. Mohn and E. Supanetz, Philos. Mag. B **78**, 629 (1998).

[44]K. R. A. Ziebeck and P. J. Webster, Philos. Mag. **34**, 973 (1976).